# PARAMETRIC OPTIMIZATION FOR AN X-RAY FREE ELECTRON LASER WITH A LASER WIGGLER

R. Bonifacio[1,3,*], N. Piovella[1,2], M.M. Cola[1,2]

[1]*INFN-Sezione di Milano, via Celoria 16, 20133 Milano, Italy*

[2]*Dipartimento di Fisica, Università degli Studi di Milano, via Celoria 16, 20133 Milano, Italy*

[3]*Centro Brasileiro de Pesquisas Fìsicas, Rio de Janeiro -Brasil*

## ABSTRACT

In this paper we optimize the experimental parameters to operate a Free Electron Laser with a laser wiggler in the Angstrom region. We show that the quantum regime of the Self Amplified Spontaneous Emission (Quantum SASE) may be reached with realistic parameters. The classical SASE regime is also discussed and compared with the quantum regime.

## INTRODUCTION

It has been shown that the quantum effects in a Free Electron Laser (FEL) are ruled by the quantum FEL parameter $\bar{\rho} = \rho(mc\gamma/\hbar k)$ [1], where $\rho$ is the classical FEL parameter [2]. The classical analysis is valid only for $\bar{\rho} \gg 1$, whereas for $\bar{\rho} < 1$ the quantum effects dominate [3,4]. In particular, in the quantum Self Amplified Spontaneous Emission (SASE) mode operation, the quantum purification of the radiation spectrum has been predicted [3,4], i.e. the broad and chaotic spectrum predicted in the classical SASE [5] and observed experimentally [6] shrinks to a very narrow spectrum when $\bar{\rho} \ll 1$.

It has been suggested that a quantum SASE FEL could be constructed using a laser wiggler [7,8] in a Compton backscattered configuration, instead of the static wiggler used in the current classical SASE experiments [6]. In a laser wiggler configuration, a low-energy electron beam back scatters the photons of a counter-propagating high power laser, with a frequency up-shifted by a factor $4\gamma^2$. The use of a laser wiggler has been discussed in the past by Gallardo et al. [9] in a classical theory.

In the following, we propose a way to optimize the experimental parameters for an X-ray FEL with a laser wiggler, showing the main differences between the parameters necessary to operate in the classical and in the quantum regimes. The analysis shows that the quantum regime appears, in general, more feasible than the classical regime for the state-of-art of electron beams and lasers technology. We stress that only in the quantum SASE regime a temporally coherent X-ray source could be realized, contrarily than in the classical SASE regime whose chaotic spectrum is temporally incoherent. Furthermore it is clear that a quantum FEL with a laser wiggler not only would be coherent but two or three orders of magnitude smaller in size (possibly table-top) and cost. The results of the present parametric study are rather encouraging with respect to a future realization of a quantum SASE X-ray FEL source.

## PARAMETRIC OPTIMIZATION

For an FEL lasing at the wavelength $\lambda_r$ with a laser wiggler with a wavelength $\lambda_L$ and wiggler parameter $a_0$, the resonance condition reads:

$$(1) \quad \gamma = \sqrt{\frac{\lambda_L (1 + a_0^2)}{4 \lambda_r}}.$$

which differs from the usual FEL resonance by a factor two, since in a laser wiggler the static wiggler period $\lambda_w$ is replaced by $\lambda_L/2$. In fact, as it is well known, the exact resonance conditions in the static and laser wiggler are

$$(2) \quad \lambda_r = \lambda_w \frac{1 - \beta_\parallel}{\beta_\parallel} = \lambda_L \frac{1 - \beta_\parallel}{1 + \beta_\parallel},$$

which implies, for $\beta_\parallel \sim 1$,

$$(3) \quad \lambda_w = \lambda_L \frac{\beta_\parallel}{1 + \beta_\parallel} \approx \frac{\lambda_L}{2}.$$

The quantum FEL parameter $\bar{\rho}$ [1] is related to the classical FEL parameter $\rho$ [2] by

$$(4) \quad \rho = \bar{\rho} \frac{\lambda_C}{\gamma \lambda_r} = \bar{\rho} \frac{2 \lambda_C}{\sqrt{\lambda_r \lambda_L (1 + a_0^2)}},$$

where $\lambda_C = h/mc = 0.024$ Å is the Compton wavelength and we used Eq.(1) to eliminate $\gamma$. In Eq.(4) $\rho$ is given by

$$\text{(5)} \qquad \rho = \frac{1}{2\gamma}\left[\frac{I}{I_A}\left(\frac{k\lambda_L a_0}{4\pi\sigma}\right)^2\right]^{1/3}.$$

where $I_A \approx 17\,\text{kA}$ is the Alfven current. Assuming as current density the peak current I divided by $2\pi\sigma^2$ or $\pi\sigma^2$, the parameter $k$ is 1 or $\sqrt{2}$ for a transversally gaussian or for a flat top shape of the current, respectively, where $\sigma$ is the beam radius. Eq. (5) is a generalization of the usual expression to a laser wiggler (see eq. (3)).

From (5) and (6) with some algebra we obtain

$$\text{(6)} \qquad I = 3.10^2 \frac{\bar{\rho}^3 \sigma^2}{k^2 \lambda_r^3 \lambda_L^2 a_0^2}$$

The units, from now on, will be $\lambda_r(\text{Å})$, $\lambda_L(\mu m)$ and $\sigma(\mu m)$. We note that the electron current is proportional to $\bar{\rho}^3$, so that, going from the quantum to the classical regime, if $\bar{\rho}$ increases for instance by a factor 10, the current increases by a factor $10^3$. This is the general reason why the use of a laser wiggler may be much more convenient in the quantum regime ($\bar{\rho} < 1$) than in the classical regime $\bar{\rho} \gg 1$.

The relation between $a_0$ and the laser power P, in agreement with [9], is

$$\text{(7)} \qquad P(TW) = \left(\frac{Ra_0}{2.4k\lambda_L}\right)^2$$

where R is the minimum radius of the laser and P is in TW and $k$ is 1 or $\sqrt{2}$ for a transversally gaussian (with beam section $2\pi R^2$) or for a flat top (with beam section $\pi R^2$) profile of the laser, respectively. From eqs. (6) and (7), we obtain the important relation between the electron current and the laser power:

$$\text{(8)} \qquad I(A) \approx 50\bar{\rho}^3 \frac{R^2\sigma^2}{k^4\lambda_L^4\lambda_r^3} \frac{1}{P(TW)}$$

As shown in ref.[7], the gain length and the cooperation length can be written in the form

$$\text{(9)} \qquad L_g = \frac{\lambda_L}{8\pi\rho}\sqrt{\frac{1+\bar{\rho}}{\bar{\rho}}}, \qquad L_c = \frac{\lambda_r}{8\pi\rho}\sqrt{\frac{1+\bar{\rho}}{\bar{\rho}}}$$

where the factor $\sqrt{1+\bar{\rho}}$ in the numerator has been added by hand to obtain the classical expression when $\bar{\rho} \gg 1$ and the quantum expression when $\bar{\rho} \ll 1$. Note that eq.(9) has a factor $8\pi$ instead of $4\pi$ in the denominator, since a laser wiggler is assumed (see eq.(3)).

Since the interaction length $L_{int}$, given by the time duration $\tau$ of the laser pulse, is approximately twice the Rayleigh range $Z_L$ of the laser, and requiring that it is larger than the gain length $L_g$, we can write the following relations:

(10) $\qquad L_{int} = c\tau \approx 2Z_L = a_1 L_g$ , $a_1 \geq 1$,

where

(11) $\qquad Z_L = \dfrac{4\pi R^2}{\lambda_L}$.

Hence the total energy of the laser pulse is given by

(12) $\qquad U = P\tau = a_1 P \dfrac{L_g}{c}$.

From Eqs. (10) and (11), we obtain the following self-consistent value of the laser rms radius at the focus:

(13) $\qquad R = \sqrt{\dfrac{a_1 L_g \lambda_L}{8\pi}}$.

Note that the power gain length is half of the value given by (9). So, if for instance $a_1=5$, then the interaction length is ten times the power gain length.

Concerning the requirement on the emittance, a very important geometrical matching condition is the following

(14) $\qquad \beta^* \equiv \dfrac{\gamma \sigma^2}{\varepsilon_n} \geq Z_L$,

where $\varepsilon_n$ is the normalized beam emittance. Eq. (14) imposes that the electron beam is contained in the laser beam, provided $\sigma \leq R$, and that the electron beam does not diverge appreciably in a Rayleigh range $Z_L$. From Eqs. (11) and (14), it follows:

(15) $\qquad \varepsilon_n \leq \varepsilon_n^{(hom)} = \dfrac{\gamma \lambda_L}{4\pi}\left(\dfrac{\sigma}{R}\right)^2$.

This is the correct condition on the emittance to be satisfied in a laser wiggler, which becomes quite restrictive when $\sigma \ll R$.

As discussed in ref. [8], the Pellegrini-Kim emittance criterium for the FEL radiation, $\varepsilon_n < \gamma\lambda_r/4\pi$, does not apply in a laser wiggler, since it would imply $\beta^* > Z_r$ where $Z_r$ is the Rayleigh range of the FEL radiation, so that the emitted radiation would get outside of the electron beam, making impossible the amplification process. To forbid this, we should reverse the criterion, i.e. $\beta^* < Z_r$ and $\varepsilon_n > \gamma \lambda_r /(4\pi)$. For simplicity we are assuming an equal radius for the radiation and the electron beams.

Furthermore, we should impose

(16) $$\sigma = \frac{R}{a_2}$$

with $a_2 > 1$ as we explained above. From eq. (15) it follows

(17) $$\varepsilon_n^{(hom)} = \frac{\gamma \lambda_L}{(4\pi)a_2^2}.$$

Up to now, all we have written is valid both in the classical and in the quantum regime. In both the cases, the condition on the energy spread is

(18) $$\frac{\Delta \gamma}{\gamma} < \Gamma,$$

where $\Gamma$ is the FEL line width. In ref.[3] we have estimated that the line width in the quantum regime is

(19) $$\Gamma = 4\rho\sqrt{\overline{\rho}} \quad \text{if} \quad \overline{\rho} < 1,$$

whereas in the classical regime it is the well known expression

(20) $$\Gamma = \rho \quad \text{if} \quad \overline{\rho} >> 1.$$

Emittance is one of the causes of the energy spread increasing. In fact, since the resonance wavelength depends on the divergence angle $\theta$, according to

(21) $$\lambda_r = \frac{\lambda_L(1 + a_0^2 + \gamma^2 \theta^2)}{4\gamma^2}, \quad \text{with} \quad 0 \le \theta \le \frac{\varepsilon_r}{\sigma},$$

we have

(22) $$\frac{\Delta \lambda}{\lambda} \approx \frac{2\Delta \gamma}{\gamma} \approx \frac{\varepsilon_n^2}{\sigma^2(1+a_0^2)} \le 2\Gamma$$

Hence, we obtain the following 'inhomogeneous' condition for emittance:

(23) $$\varepsilon_n \le \varepsilon_n^{(in\,hom)} = \sigma\sqrt{2\Gamma(1+a_0^2)}$$

which, using eq. (9), is equivalent to

(24) $$\varepsilon_n \le \alpha \frac{\gamma \lambda_r}{4\pi} \sqrt{\frac{Z_r}{L_g}}$$

where $\alpha = 2$ for the classical case and $\alpha = 4$ for the quantum case, in agreement with refs.[8,10].

We remark that the inequality (15) must be strictly satisfied, otherwise the FEL action is destroyed. The inequality (23) or (24) arises from an inhomogeneous broadening of the resonance which may reduce the emission deteriorating the gain, since only the electrons whose θ is small enough will participate to the radiation process [11].

Another cause which may contribute to the broadening of the resonance is the intensity fluctuations in the laser wiggler, i.e. the fluctuations in the wiggler parameter $a_0$. Using Eq. (1) and imposing (18) we obtain, very simply,

(25) $$\frac{\Delta a_0}{a_0} \leq \frac{1+a_0^2}{a_0^2}\Gamma.$$

Finally, the peak power in an FEL is given by [2]

(26) $$P_r = P_{beam}(\rho|A|^2)$$

where A is the dimensionless field amplitude in the 'universal scaling' and $P_{beam} = (I/e)mc^2\gamma$ is the beam power. Classically, at saturation, $|A|^2 \approx 1$ [2], so that eq. (26) can be written

(27) $$P_{sat} \approx \rho P_{beam} \approx (I/e)\hbar\omega\bar{\rho}.$$

Eq.(27) shows that $\bar{\rho}$ is the average number of the emitted photons per electron at saturation. In the quantum regime $\bar{\rho} < 1$ and at saturation [3], $|A|^2 \approx 1/\bar{\rho}$ and eq.(26) yields

(28) $$\bar{P}_{sat} \approx \frac{\rho}{\bar{\rho}}P_{beam} \approx \frac{\hbar\omega}{mc^2\gamma}P_{beam} \approx (I/e)\hbar\omega$$

Note that the power in the quantum regime is larger than the one predicted in the classical theory. The meaning of Eq. (28) is transparent: in the quantum regime each electron emits a single photon. In conclusion, the number of emitted photons in the classical and in the quantum regimes is respectively

(29) $$N_{ph} = \frac{Q}{e}\hbar\omega$$

and

(30) $$\bar{N}_{ph} = \frac{Q}{e}\hbar\omega\bar{\rho}.$$

## CLASSICAL VERSUS QUANTUM REGIME

In order to discuss some specific example for the classical and the quantum regimes, we take as independent the following system of six parameters: $\bar{\rho}$, $\lambda_r$(Å), $\lambda_L$(μm), $a_0$ (the wiggler parameter), $a_1$ (the number of amplitude gain lengths in the interaction region $2Z_L$) and $a_2$ (the ratio between the laser and the radiation beam radius at the focal point). The other parameters are deduced self-consistently as follows. Using eq. (1) we deduce $\gamma$. With eq. (9) we calculate $L_g$, with eq. (13) we deduce R and than $\sigma$ from (16). Introducing these values in eqs. (6), (7) and (12) we calculate the current I, the power P and the total energy U. Finally we calculate the limit values on emittance from (17) and (23) and the number of photons from eqs. (29) and (30), for a given beam charge Q. In the table 1 we report the results of the optimization with $\lambda_r$=2Å, $\lambda_L$=0.8 µm, $a_1$=5 and $a_2$=2, both for a quantum case, with $\bar{\rho}=0.2$ and $a_0$=0.1, and for a classical case, with $\bar{\rho}=2$ and $a_0$=0.8. Furthermore, we assume Q=1 nC.

| | | |
|---|---|---|
| $\bar{\rho}$ | 0.2 | 2 |
| $a_0$ | 0.1 | 0.8 |
| $\rho$ | $7.55 \cdot 10^{-5}$ | $5.93 \cdot 10^{-4}$ |
| $\gamma$ | 32 | 40 |
| $\Gamma$ | $1.35 \cdot 10^{-4}$ | $6 \cdot 10^{-4}$ |
| $\Delta a_0 / a_0$ | $1.36 \cdot 10^{-2}$ | $1.5 \cdot 10^{-3}$ |
| $L_g$ (mm) | 1 | 0.06 |
| $Z_L$ (mm) | 2.6 | 0.17 |
| R (µm) | 12.9 | 3.25 |
| $\sigma$ (µm) | 6.4 | 1.6 |
| P (TW) | 0.23 | 0.92 |
| Energy (J) | 3.9 | 1.0 |
| $\tau$ (ps) | 17.1 | 1.1 |
| I (A) | 990 | 985 |
| Photons' number | $6.2 \cdot 10^9$ | $12.4 \cdot 10^9$ |
| $\varepsilon_n^{(hom)}$ | 0.5 | 0.6 |
| $\varepsilon_n^{(in\,hom)}$ | 0.11 | 0.07 |

Table 1

From our results it appears that in the quantum regime the gain length and the laser Rayleigh range are appreciably longer than in the classical regime, so that in the quantum regime a longer duration time and a larger energy of the laser are required. The current in both cases is of the order of 1kA, however the requirement on the current density in the classical case is an order of magnitude larger since an electron beam radius of 1.6 micron is very small. We note also a more stringent condition for the inhomogeneous emittance and the laser fluctuations in the classical case. Therefore, on the basis of this example, we would conclude that it is easier to operate in the quantum regime. We stress that in the quantum regime the emitted radiation has the important property of high temporal coherence with no spiking, whereas for the classical regime, with $\bar{\rho}=2$, one would have nearly 3000 random spikes. This is the fundamental difference between the two regimes. A 3D quantum model for a FEL with a laser wiggler will be discussed elsewhere.